\documentstyle[preprint,aps,epsfig]{revtex}

\def\be{\begin{equation}}
\def\ee{\end{equation}}
\def\ep{\epsilon}

\begin{document}
\draft

\title{First-principles approach to dielectric response of \\ 
 graded spherical particles} 
\author{L. Dong$^1$, G. Q. Gu$^{1,2}$, K. W. Yu$^1$}
\address{$^1$Department of Physics, The Chinese University of Hong Kong,
 Shatin, NT, Hong Kong}
\address{$^2$College of Information Science and Technology, 
 East China Normal University, \\ Shanghai 200 062, China}
\maketitle

\begin{abstract}
We have studied the effective response of composites of spherical particles 
each having a dielectric profile which varies along the radius of the 
particles. We developed a first-principles approach to compute the dipole 
moment of the individual spherical particle and hence the effective 
dielectric response of a dilute suspension. 
The approach has been applied to two model dielectric profiles, 
for which exact solutions are available.
Moreover, we used the exact results to validate the results from the 
differential effective dipole approximation, recently developed to treat 
graded spherical particles of an arbitrary dielectric profile. 
Excellent agreement between the two approaches were obtained.
While the focus of this work has been on dielectric responses, 
the approach is equally applicable to analogous systems such as the 
conductivity and elastic problems.

\end{abstract}
\vskip 5mm 
\pacs{PACS Number(s): 77.22.E, 77.84.L, 42.79.R}

\section{Introduction}

Functionally graded materials (FGM) are inhomogeneous materials with 
spatially varying material properties. These materials have received 
considerable attention as one of the advanced inhomogeneous composite 
materials in various engineering applications since first being reported 
in the 1980s \cite{Yamanouchi}. 
The change in the composition induces material and microstructure 
gradients, and makes the FGM very different in behavior from the 
homogeneous materials and conventional composite materials 
\cite{Yamanouchi,Holt}. 
These materials can be tailored in their materials properties via the 
design of the gradients. For mechanical properties, 
the main advantages of a graded material profile range from the improved 
bonding strength, toughness, to wear and corrosion resistance. 
Other benefits include the reduced residual and thermal barrier coatings 
of high temperature components in gas turbines, the surface hardening for 
tribological protection and graded interlayers used in multilayered 
microelectronic and optoelectronic components 
\cite{Yamanouchi,Holt,Ilschner}.

Over the past few years, there have been a number of attempts, both 
analytical and experimental, to study the responses of FGM to mechanical 
\cite{Atkinson,Delale,Erdogan,Chen,PGu}, thermal \cite{Jin,Jin1,Noda} and 
electric \cite{Zhu,Scanchez} loads and for different microstructure in 
various systems. Nevertheless, the responses of composites of FGM 
inclusion, which is formed when graded inclusions are suspended randomly 
in a host medium, should be more useful and interesting. 
Various different attempts have been made to treat the composite materials 
of homogeneous inclusions \cite{Jackson} as well as multi-shell inclusions 
\cite{Gu1,Fuhr,Arnold,Chan};
there exist many methods to study the effective properties of composite 
materials in the literature. 
However, these established theories for homogeneous inclusions cannot be 
applied. It is thus necessary to develop a new theory to study the 
effective properties of graded composite materials under externally applied 
fields. In this paper, we will develop a first-principles approach for 
calculating the effective response of graded composite materials.

The paper is organized as follows. In the next section, we present the 
formalism. We will solve the field equations analytically for graded 
spherical inclusions. In sections III, we obtain the exact solution for a 
power-law profile, and in section IV, for a linear profile. 
In section V, we use the exact solution to validate a recently proposed 
approximate theory. Discussion and conclusion will be given.


\section{Formalism}

In this section, we will focus on the dielectric response as an example. 
The formalism can equally be applied to analogous systems like the 
conductivity and elastic problems.
More precisely, for a composite of graded spherical inclusions, we will 
find its response to an externally applied electric field. 
The formalism involves two major parts: firstly we will calculate the 
local electric field distribution in a graded spherical inclusion and 
then the induced dipole moment of the inclusion.
The two-dimensional composite of graded cylindrical inclusions has been 
discussed in \cite{Gu}. In this work, we will extend the analysis
to the more realistic case of three-dimensional composites of graded 
spherical inclusions. We will study two model dielectric profiles, 
namely, the power-law and linear profiles for which exact solutions of 
the local electric field and the dipole moment can be obtained.
 
We consider a graded spherical inclusion of radius $a$ suspended in a 
homogeneous host medium, subjected to a uniform electric field 
$\vec{E}_0$ along the $z$-axis. 
We will consider a low concentration of suspended inclusions. 
Thus, the interaction among the inclusions can be neglected and the 
effective dielectric properties of the composite can be obtained from the 
responses of a single inclusion under an effective electric field $\bar{E}$. 
For dielectric response, the constitutive relation of a graded spherical 
inclusion reads
\be
\vec{D}=\ep_{i}(r)\vec{E}_{i}.
\ee
where $\ep_{i}(r)$ is the dielectric profile of the graded spherical 
inclusion. The relation for the host medium is 
\be
\vec{D}=\ep_{m}\vec{E}_{m}.
\ee
where $\ep_{m}$ is the dielectric constant of the host medium. 
The Maxwell's equations read
\be
\vec{\nabla} \cdot \vec{D}=0,
\label{gradient}
\ee
and
\be
\vec{\nabla} \times \vec{E}=0.
\ee
To this end, $\vec{E}$ is the gradient of a scalar potential $\Phi$:
\be
\vec{E}=-\vec{\nabla}\Phi.
\label{pot}
\ee
Eq.(\ref{gradient}) and Eq.(\ref{pot}) can be combined into a partial 
differential equation: 
\be
\vec{\nabla} \cdot (\ep(r) \vec{\nabla} \Phi) = 0.
\ee
We will normalize the dielectric profile to the dielectric constant of 
the host medium $\ep_{m}$ for convenience. Without loss of generality, we 
may also let $a=1$.

In spherical coordinates, the electric potential $\Phi$ satisfies
\be
\frac{1}{r^2}\frac{\partial}{\partial r}
\left(r^2\ep(r) \frac{\partial \Phi}{\partial r}\right)
+\frac{1}{r^2\sin\theta}\frac{\partial}{\partial \theta}
\left(\sin\theta \ep(r) \frac{\partial \Phi}{\partial r}\right)
+\frac{1}{r^2\sin^2\theta}\frac{\partial}{\partial \varphi}
\left(\ep(r)\frac{\partial \Phi}{\partial \varphi}\right)=0,
\label{comm}
\ee
where $\ep(r)=\ep_{i}(r)/\ep_{m}$, $\ep_{i}(r)$ is the dielectric 
constant of the inclusion and $\ep_{m}$ is the dielectric constant of the 
host medium.

We consider the applied electric field along the $z$-axis, thus $\Phi$ 
is independent of the angle $\varphi$. If we write $\Phi = R(r)\Theta(\theta)$,
after a separation of variables, we obtain the ordinary differential 
equation for the radial function $R(r)$,
\be
\frac{d}{dr}\left(r^2\frac{dR}{dr}\right)
+\frac{r^2}{\ep(r)}\frac{d\ep(r)}{dr}\frac{dR}{dr}-n(n+1)R=0.
\label{general}
\ee
where $n$ is an integer, $\ep(r)$ is a dimensionless dielectric constant, 
while $\ep(r)=\ep_{i}(r)/\ep_{m}$ in the inclusion, and $\ep(r)=1$ in the 
host medium.

The potential can be obtained by solving Eq.(\ref{general}). 
In the dilute limit, the dipole moment of the graded spherical inclusion 
can be derived. We take the average of the operator 
$\vec{D}-\ep_{m}\vec{E}$ in the whole volume of the composite 
material, then
\be
\frac{1}{V}\int_{V} (\vec{D}-\ep_{m}\vec{E}) dV
=\bar{D}-\ep_{m}\bar{E},
\label{operator}
\ee
where $V$ is the volume of the whole composite material, $\bar{A}$ denotes 
the average of an operator $A$ in the composite material.
The integrand vanishes in the host medium, and thus Eq.(\ref{operator}) 
becomes 
\be
\frac{1}{V}\int_{\Omega_{i}} (\vec{D}-\ep_{m}\vec{E}) dV
=\bar{D}-\ep_{m}\bar{E},
\ee
where $\Omega_{i}$ is the region occupied by the inclusion.

Now, we can define the effective constitutive relation for the composite 
material:
\be
\bar{D}=\ep_{e}\bar{E},
\ee
where $\ep_{e}$ is the effective dielectric constant of the composite 
material. Thus 
\be
\frac{1}{V}\int_{\Omega_{i}} (\ep_{i}-\ep_{m}) \vec{E} dV
=(\ep_{e}-\ep_{m})\bar{E}.
\label{equation}
\ee
Eq.(\ref{equation}) gives the polarization of the composite and it can be 
used to calculate the effective dielectric properties of the composite 
material at a low particle concentration.


\section{Exact solution for a power-law profile}

The dielectric constant in the particle is taken as a power-law function 
of the radius.
In this case, $\ep(r)=c_{k}r^k$, with $k \ge 0$ where $0\le r \le1$.
Then the radial equation becomes 
\be
\frac{d^2 R}{dr^{2}}+\frac{k+2}{r}\frac{dR}{dr}-\frac{n(n+1)}{r^2}=0.
\label{Re}
\ee
As this is a homogeneous equation, it admits a power-law solution,
\be
R(r)=r^{s}.
\ee
\label{R(r)}
Substituting it into Eq.(\ref{Re}), we obtain the equation
\be
s^{2}+s(k+1)-n(n+1)=0.
\ee
The solution to this equation is
\be
s^{k}=\frac{1}{2}\left[-(k+1)\pm\sqrt{(k+1)^{2}+4n(n+1)} \right].
\ee

The potentials in the inclusion and the host medium are given, 
respectively, by
\begin{eqnarray}
\Phi_{i}(r,\theta)&=&A_{0}+\sum_{n=0}^{\infty} \left[A_{n} 
r^{s_{+}^{k}(n)}+B_{n}r^{s_{-}^{k}(n)} \right]P_{n}(\cos\theta),\nonumber\\
\Phi_{m}(r,\theta)&=&C_{0}+\sum_{n=0}^{\infty} \left[C_{n}
r^{n}+D_{n}r^{-(n+1)} \right]P_{n}(\cos\theta).
\label{potential}
\end{eqnarray}

As $r\to \infty$, the potential is given by the far field
\be 
\Phi_{m}=-E_{0}r\cos\theta,
\ee
while $r\to 0$, $\Phi_{i}$ attains a finite value. Thus
\begin{eqnarray}
\Phi_{i}(r,\theta)&=& A_{1} r^{s_{+}^{k}(1)} P_{1}(\cos\theta),
\nonumber\\
\Phi_{m}(r,\theta)&=&-E_{0}r\cos\theta
+\frac{D_{1}}{r^{2}}P_{1}(\cos\theta).
\end{eqnarray}
Meanwhile, the potential functions should satisfy the boundary conditions,
as follow: 
\begin{eqnarray}
\Phi_{i}(r,\theta)\left|_{r=1} \right.&=&\Phi_{m}(r,\theta)\left|_{r=1} 
\right.,\nonumber\\
\ep(r)\frac{\partial \Phi_{m}(r,\theta)}{\partial r}\left|_{r=1} 
\right.&=&\frac{\partial 
\Phi_{i}(r,\theta)}{\partial r}\left|_{r=1} \right..
\label{bc}
\end{eqnarray}

We obtain:
\be
-E_{0}+D_{1}=A_{1},\ \ \ 
-E_{0}-2D_{1}=c_{k}A_{1}s_{+}^{k}(1).
\ee

It is not difficult to solve these equations and obtain the coefficients,
\be
A_{1}=\frac{-3}{sc_{k}+2}E_{0},\ \ \
D_{1}=\frac{sc_{k}-1}{sc_{k}+2}E_{0}. 
\ee
where $s=s_{+}^{k}(1)=\frac{1}{2}\left[\sqrt{(k+1)^{2}+8} - (k+1)\right]$. 
Then the potentials are given by
\begin{eqnarray}
\Phi_{m}(r,\theta)&=&-E_{0}r\cos\theta+\frac{D_{1}}{r^{2}}\cos\theta, 
\nonumber\\ 
\Phi_{i}(r,\theta)&=&A_{1}r^{s}\cos\theta.
\end{eqnarray}

The electric field $E_{i}$ along the $z$-axis in the inclusion can be 
derived from the potential $\Phi_{i}$,
\be
E_{iz}=-A_{1}r^{s-1}\left[(s-1)\cos^{2}\theta +1 \right].
\label{E}
\ee

In the dilute limit, the polarization can be calculated from the 
following equation: 
\be
4 \pi \ep_{m}E_{0} b=\int_{\Omega_{i}} 
(\ep_{i}(r)-\ep_{m})E_{iz}(r,\theta) dV,
\ee
where $b$ is the dipole factor, which measures the degree of polarization 
of the graded spherical inclusion against the host medium, and $\Omega_{i}$ 
is the region occupied by the inclusion. After performing the integral, 
we obtain 
\be
b=\frac{s+2}{sc_{k}+2}\left(\frac{c_{k}}{k+s+2}-\frac{1}{s+2} \right).
\label{b}
\ee
When $k=0$, $\ep(r)=c_{k}$ is a constant, then
\be
s=1
\ee
and
\be
b=\frac{c_{k}-1}{c_{k}+2}
\ee
This is just the dipole factor in the homogeneous case.


\section{Linear profile with a small slope}

In this section, we consider $\ep(r)$ as a linear function of the radius. 
This is a simple graded material that can be made easily. 
Moreover, other materials can be simulated using piecewise linear functions. 
For the inclusion, 
\be
\ep(r)=d+cr.
\ee
According to Eq.(\ref{general}), the radial function for the inclusion 
satisfies the following equation: 
\be
\frac{d^{2}R}{d\hat{r}^{2}}+\left(\frac{2}{\hat{r}}
+\frac{1}{\hat{r}+1}\right)\frac{dR}{d\hat{r}}
-\frac{n(n+1)}{\hat{r^{2}}}R=0.
\label{lineareq}
\ee
where $r=\xi \hat{r}$, and $\xi=d/c$. 
We consider the case with a small slope, that is, 
$\vert \xi \vert=\vert d/c \vert >1$, we can express the solution in a 
power series. The power series solution is: 
\be
f_{n}(\hat{r})=\sum_{k=0}^{\infty}C_{k}^{n}\hat{r}^{k+\rho}.
\ee
Then the radial equation [Eq.(\ref{lineareq})] becomes
\begin{eqnarray}
\sum_{k=0}^{\infty}C_{k}^{n}\left[(k+\rho)(k+\rho-1)+2(k+\rho)-n(n+1)
\right]\hat{r}^{k+\rho-2}\nonumber\\
+ \sum_{k=0}^{\infty}C_{k}^{n}\left[(k+\rho)(k+\rho-1)
+3(k+\rho)-n(n+1)\right]\hat{r}^{k+\rho-1}=0.
\label{relation}
\end{eqnarray}
The coefficient of each term should vanish, and the lowest term satisfies: 
\be
\left[\rho(\rho-1)+2\rho-n(n+1) \right]C_{0}^{n}=0.
\ee
This equation is for the characteristic equation of the differential 
equation [Eq.(\ref{lineareq})]. Solving it, we obtain
\be 
\rho= n\ \ {\rm or}\ \ -(n+1).
\ee
Similarly, the recursion relation can also be found from 
Eq.(\ref{relation}), 
\be
C_{k+1}^{n}=-\frac{(k+n)(k+n+2)-n(n+1)}{(k+n+1)(k+n+2)-n(n+1)}C_{k}^{n}.
\ee
This power series solution is absolutely convergent in the region 
$0\leq \hat{r}<1$. Thus, we can use this solution to study the response of 
composite material to the external field, as long as $\vert d/c \vert >1$.

Consider the properties of the potential at infinity and at the origin, 
the potentials are given by
\begin{eqnarray}
\Phi_{m}(r,\theta)&=&-E_{0}r\cos\theta+\frac{D_{1}}{r^{2}}\cos\theta, 
\nonumber\\
\Phi_{i}(r,\theta)&=&A_{1}\sum_{k=0}^{\infty}C_{k}^{1}
(\frac{c}{d}r)^{k+1}\cos\theta.
\end{eqnarray}
Moreover, from the boundary conditions, we obtain the coefficients 
$A_{1}$ and $D_{1}$,
\be
A_{1}=\frac{-3E_{0}}{(d+c)\upsilon_{2}+2\upsilon_{1}}, \ \ \
D_{1}=\frac{(d+c)\upsilon_{2}-\upsilon_{1}}{(d+c)\upsilon_{2}
+2\upsilon_{1}}E_{0}.
\ee
where
\be
\upsilon_{1}=\sum_{k=0}^{\infty}C_{k}^{1}\left(\frac{c}{d} 
\right)^{k+1},\ \ \
\upsilon_{2}=\sum_{k=0}^{\infty}C_{k}^{1}(k+1)\left(\frac{c}{d} 
\right)^{k+1}. 
\ee
Then the $z$-component of the electric field can also be calculated
\be
E_{iz}=-A_{1}\sum_{k=0}^{\infty}C_{k}^{1}\left(\frac{c}{d} 
\right)^{k+1}r^{k}(k\cos^{2}\theta+1). 
\ee
Thus we obtain the dipole factor
\be
b=\frac{1}{(d+c)\upsilon_{2}+2\upsilon_{1}}
\left[(b-1)\upsilon_{1}+c\upsilon_{3} \right].
\ee
where
\be
\upsilon_{3}=\sum_{k=0}^{\infty}C_{k}^{1}\left(\frac{c}{d}\right)^{k+1}
\frac{k+3}{k+4}. 
\ee
When $c=0$, $\ep(r)$ is a constant, Eq.(40) reduces to the homogeneous 
case. Thus, when $c\to 0$, the dipole factor is given by 
\be
b=\frac{d-1}{d+2}=\frac{\ep_{i}-\ep_m}{\ep_{i}+2\ep_m}.
\ee
This result is the same with that in the homogeneous case.


\section{Comparison with differential effective dipole approximation}
 
In this section, we compare the exact results with the differential 
effective dipole approximation (DEDA) \cite{Yu}, 
recently developed to treat arbitrary graded profiles. 
We start with the differential equation for the dipole factor \cite{Yu}:
\be
{db\over dr} = -{1\over 3r\ep_{m}\ep(r)}[(1 + 2b)\ep_{m} - (1 - b)\ep(r)]
[(1 + 2b)\ep_{m} + 2(1 - b)\ep(r)],
\ee
where $\ep_{m}=1$.
Thus the dipole factor of a graded spherical inclusion can be calculated 
by solving the above differential equation with a given graded profile 
$\ep(r)$. The nonlinear first-order differential equation can be 
integrated, if we are given the graded profile $\ep(r)$ and the initial 
condition $b(r=0)$. 

We have evaluated DEDA for for two model graded profiles:
(a) power-law profile $\ep(r) = C r^n$, and 
(b) linear profile $\ep(r) = A + B r$.
Note that we have changed notations slightly to agree with those of 
Ref.\cite{Yu}.
The numerical integration has been done by the fourth-order Runge-Kutta 
algorithm with a step size $\delta r=0.01$.
In Fig.1(a), we plot the dipole factor $b$ versus $C$ for various index 
$n>0$. It is clear that $b$ increases monotonically as the dielectric 
contrast $C$ increases, while it decreases with the index $n$.
It is attributed to the fact that the average dielectric constant decreases 
as $n$ increases.
Similarly, in Fig.1(b) we plot $b$ versus $A$ for various slope $B$. 
We obtained similar behavior as in Fig.1(a).

It is instructive to compare the exact results with the DEDA results.
In Fig.1, we compared the exact results with the DEDA results for a 
spherical inclusion with a power-law profile and a linear profile. 
The agreement is excellent.
Note that exact results are unavailable for $A<B$ in the linear profile.
Thus, we would say the DEDA is a very good approximation for graded 
spherical inclusions.


\section{Discussion and conclusion}

Here a few comments are in order.
In this work, we have developed a first-principles approach to compute 
the dipole moment of the individual spherical inclusion and hence the 
effective dielectric response of a dilute suspension. 
The approach has been applied to two model dielectric profiles, 
for which exact solutions are available.
Moreover, we used the exact results to validate the results from the 
differential effective dipole approximation, which is valid for graded 
spherical inclusions of an arbitrary dielectric profile. 
Excellent agreement between the two approaches were obtained.
Note that an exact solution is very few in composite research and to have 
one yields much insight. It is thus worth spending time on finding one. 
To this end, it is also instructive to obtain the analytic result for the 
piecewise linear profiles \cite{Gu}.

It is instructive to extend the present approach to nonlinear graded 
composites. The introduction of a graded dielectric profile in nonlinear 
composites provides an extra dimension to tune the enhanced nonlinear 
response. For instance, a graded interfacial layer on the spherical 
inclusions may help to control the local field fluctuations, and hence 
the nonlinear response. The perturbation approach \cite{Gu-Yu} as well as 
the variational approach \cite{Yu-Gu} are just suitable for this extension. 

Our approach can be applied to biological cells, as the interior of 
biological cells must be inhomogeneous in nature and can be treated as a 
graded material. 
To this end, Freyria et. al. \cite{Freyria} observed a graded cell 
response when they studied the cell-implant interactions experimentally. 
Thus the complex dielectric function can be modeled to vary continuouly 
along the radius of the cells, namely, the conductivity can change 
rapidly near the boundary of cells and a power-law profile prevails, 
while the dielectric function may vary only slightly and thus a linear 
profile suffices. Work is in progress along this direction \cite{Huang}.

\section*{Acknowledgments}
This work was supported by the RGC Earmarked Grant under project number 
CUHK 4245/01P. G.Q.G. acknowledges his research grant here.
We thank J. P. Huang for his assistance in the numerical solutions and 
plotting figures.

\begin{figure}[h]
\caption{The dipole factor $b$ plotted for two model profiles: 
(a) versus $C$ for various index $n$ in the power-law profile, and 
(b) versus $A$ for various slope $B$ in the linear profile.}
\end{figure}

\centerline{\epsfig{file=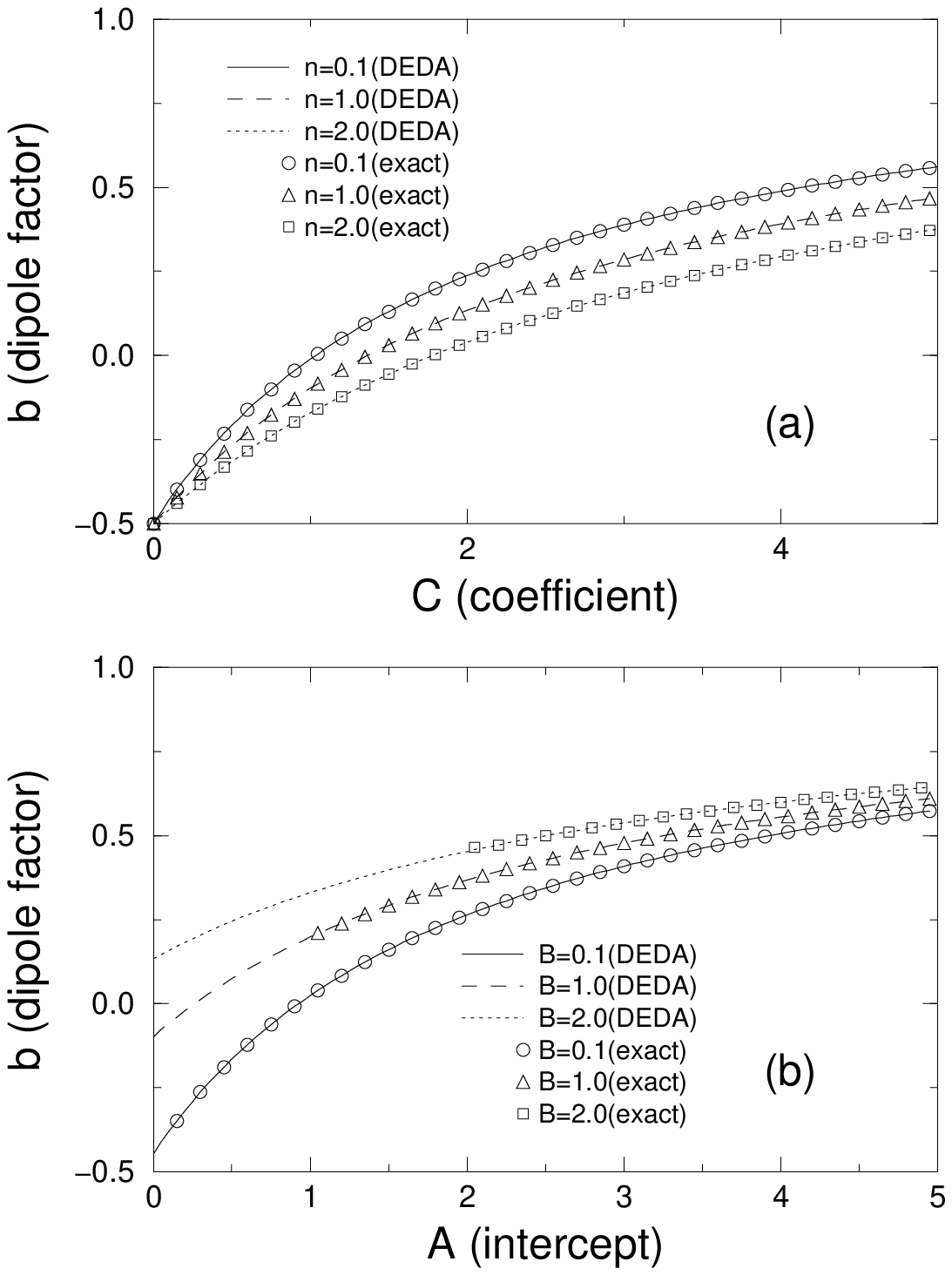,width=360pt}}
\centerline{Fig.1/Dong, Gu, Yu}

\end{document}